\documentclass[a4paper]{llncs}

\usepackage{amssymb,amsmath}
\usepackage{graphicx}
\usepackage{oz}
\usepackage{enumerate}

\usepackage{url}
\usepackage{wrapfig}

\begin{document}

\mainmatter  

\title{A macro-level model for investigating the effect of directional bias on network coverage}


\author{Graeme Smith$^1$
\and J.W. Sanders$^{2,3}$ \and Qin Li$^1$}
%

\institute{$^1$School of Information Technology and Electrical Engineering\\
The University of Queensland, Australia\\
$^2$African Institute for Mathematical Science, South Africa\\
$^3$Department of Mathematical Sciences, Stellenbosch University, South Africa\\
}

\maketitle

\begin{abstract}
Random walks have been proposed as a simple method of efficiently searching, or disseminating information throughout, communication and sensor networks. In nature, animals (such as ants) tend to follow {\em correlated\/} random walks, i.e., random walks that are biased towards their current heading. In this paper, we investigate whether or not complementing random walks with directional bias can decrease the expected discovery and coverage times in networks. 

To do so, we  develop a macro-level model of a directionally biased random walk based on Markov chains. By focussing on regular, connected networks, the model allows us to efficiently calculate expected coverage times for different network sizes and biases. Our analysis shows that directional bias can  significantly reduce coverage time, but only when the bias is below a certain value which is dependent on the network size. 
\end{abstract}

\section{Introduction}
\label{sec:intro}

The concept of a random walk was introduced over a century ago by Pearson \cite{pea05} and has been studied extensively since then \cite{dvo51,bru91,cas96}. Recently, random walks have been proposed for searching, or disseminating information throughout, communications and sensor networks where the network's structure is dynamic, or for other reasons unknown \cite{bar06,dol06,avi04,sad05}.  They are ideal for this purpose as they require no support information like routing tables at nodes \cite{avi06} --- the concept of a random walk being for the agent performing the walk to move randomly to {\em any\/} connected node. 

The efficiency of random-walk-based algorithms can be measured in terms of the average number of steps the agent requires to cover every node in the network (and hence be guaranteed to find the target node in the case of search algorithms). This is referred to as the {\em coverage time\/} under the assumption that the agent takes one step per time unit. Obviously, improving the coverage time for algorithms is an important goal. 

One straightforward approach to this is to have multiple agents \cite{alo11}. For some algorithms, such an approach is made even more effective when stigmergy is employed \cite{par97,bon98,gho08}, i.e., agents leave information for other agents directing them to their goals. Such an approach, inspired by the way ants leave trails of pheromones directing other ants to food \cite{cam01}, is only useful when target nodes need to be visited by more than one agent. This is not always the case. More importantly, stigmergy is effective in directing agents only once a target node has been found. The time for the first agent to find a target node is not reduced. This can only be done by considering the agent's `movement model'. 

For this reason it has been suggested that random walks should be constrained, e.g., to prevent an agent returning to its last visited node, or to direct an agent to parts of the network where relatively few nodes have been visited \cite{lim07}. We take a similar approach in this paper. We base our movement model on that observed in nature. Many models used by biologists to describe the movement of ants and other animals are based on {\em correlated\/} random walks, i.e., random walks which are biased to the animal's current direction \cite{kar83,bov88,mcc89}. Based on our own observations of ants, we also investigate including a small probability of a non-biased step at any time to model occasional random direction changes. 

To the best of our knowledge, directionally biased walks in networks have been investigated by only one other group of researchers. Fink et al.\ \cite{fin12} look at the application of directional bias in a cyber-security system in which suspect malicious nodes must be visited by multiple agents. They compare coverage times for directional bias, with those for pure random walks, and random walks with stigmergy. Their conclusion is that directionally biased walks are more efficient even than random walks with stigmergy. This conclusion, however, is based on micro-level simulation, i.e., direct simulation of agents taking steps, for a single network size and bias. It cannot be generalised to arbitrary network size or bias. 

The micro-level simulation approach of Fink et al.\ requires coverage times to be calculated as the average of multiple runs. They performed 500 simulation runs for each movement model. Such an approach is impractical for a deeper investigation of the effect of directional bias which considers various network sizes and biases. For that reason, in this paper we develop a more abstract, macro-level model of a directionally biased walk. It builds on the work of Mian et al.\ \cite{mia10} for random walks, describes the directionally biased walk in terms of a Markov chain \cite{nor98} and allows us to calculate the coverage time for a given network size and bias directly. Although certain special cases have analytic solutions, we have found this model to be helpful for a general approach to calculating coverage time.

We begin in Section~\ref{sec:prob} by describing the concept of a random walk, and our notion of directional bias in more detail. In Section~\ref{sec:macro} we present the Markov-chain model of a directionally biased walk on a network and show how it can be used to calculate the expected coverage time. In Section~\ref{sec:res} we present and discuss the results of applying our model to the calculation of coverage times on a range of network sizes and biases. We conclude in Section~\ref{sec:conc}.

\section{Directionally biased walks}
\label{sec:prob}

Random walks have been studied in 1-dimensional, 2-dimensional and multi-dimensional spaces. Many of the results from 2-dimensional walks are applicable to communications and sensor networks which are commonly modelled as connected graphs. In particular, it is known that with probability 1 a random walk will cover every node of a connected graph \cite{ald89} and a number of approaches for calculating the coverage time have been proposed \cite{ald91,dem04,lim07,mia10}.

\begin{figure}[t]
  \begin{center}
 \includegraphics[scale=0.65]{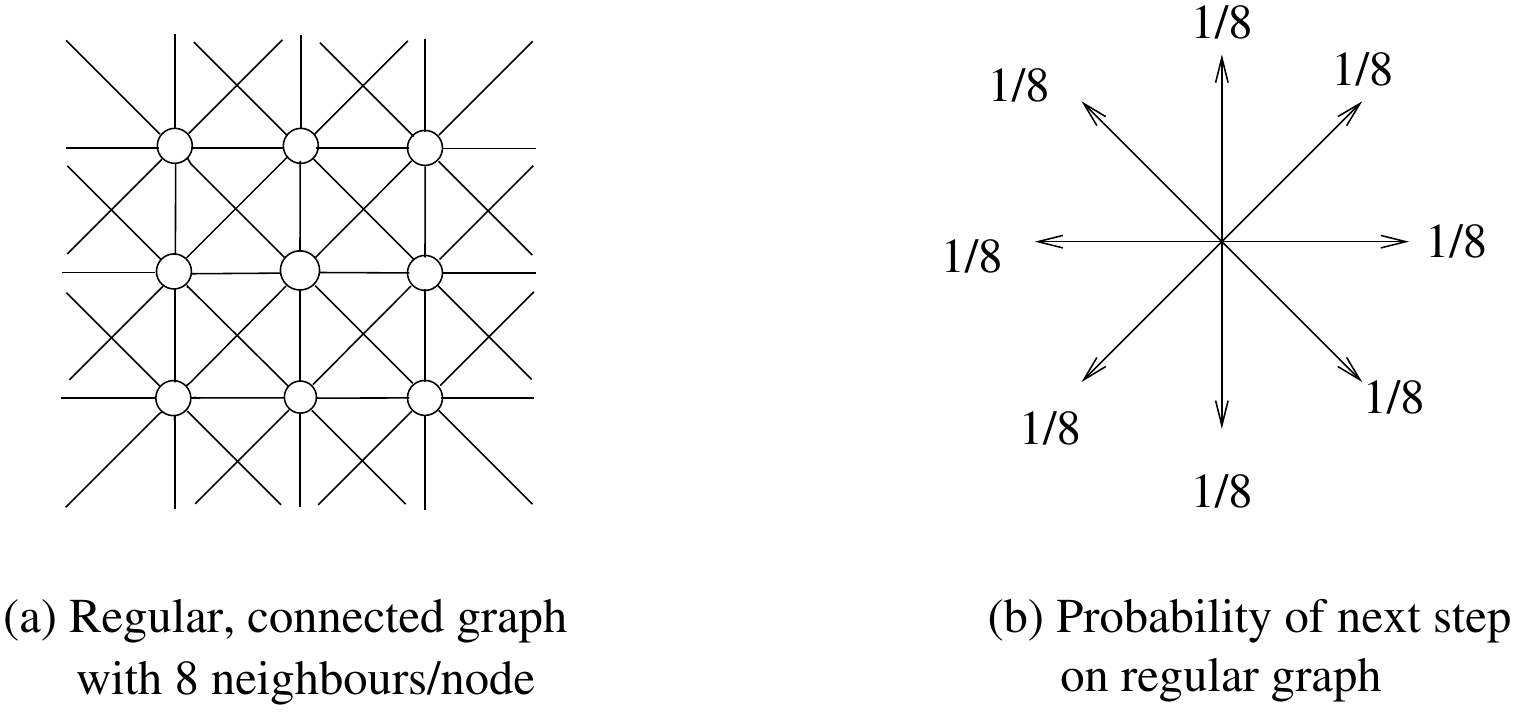}
  \end{center}
  \caption{Random walk on a regular, connected graph where each node has 8 neighbours.}
   \label{fig:random}
\end{figure}

The investigation in this paper focusses on regular, connected graphs where each node has exactly 8 neighbours (see Fig.~\ref{fig:random}(a) and, for the probability of the next step in a random walk over such a graph, Fig.~\ref{fig:random}(b)). Furthermore, to allow our graphs and hence networks to be finite, we wrap the north and south edges and the east and west edges to form a torus. Our aim is to provide a deeper analysis of directional bias than in the recent literature which also investigates the notion on regular toroidal graphs \cite{fin12}. While the results do not apply directly to arbitrary irregular networks, they do apply to random geometric graphs which are often used to model ad hoc sensor networks. Approaches using regular toroidal graphs to determine  coverage time on random geometric graphs include that of Lima and Barros \cite{lim07} and Mian, Beraldi and Baldoni \cite{mia10}.

%

\begin{figure}[t]
  \begin{center}
 \includegraphics[scale=0.22]{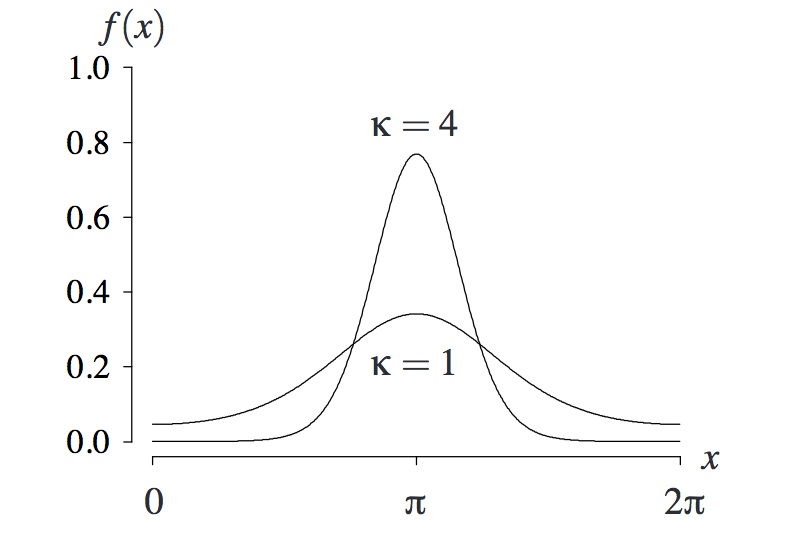}
  \end{center}
  \caption{The von Mises distribution about $\pi$ radians for $\kappa=1$ and $\kappa=4$.}
   \label{fig:vm}
\end{figure}

For modelling directional bias in nature, biologists typically use the von Mises distribution \cite{cri91}. The von Mises distribution is a continuous angular function with a parameter $\kappa$ which affects heading bias (see Figure~\ref{fig:vm}).

\begin{figure}[t]
  \begin{center}
 \includegraphics[scale=0.65]{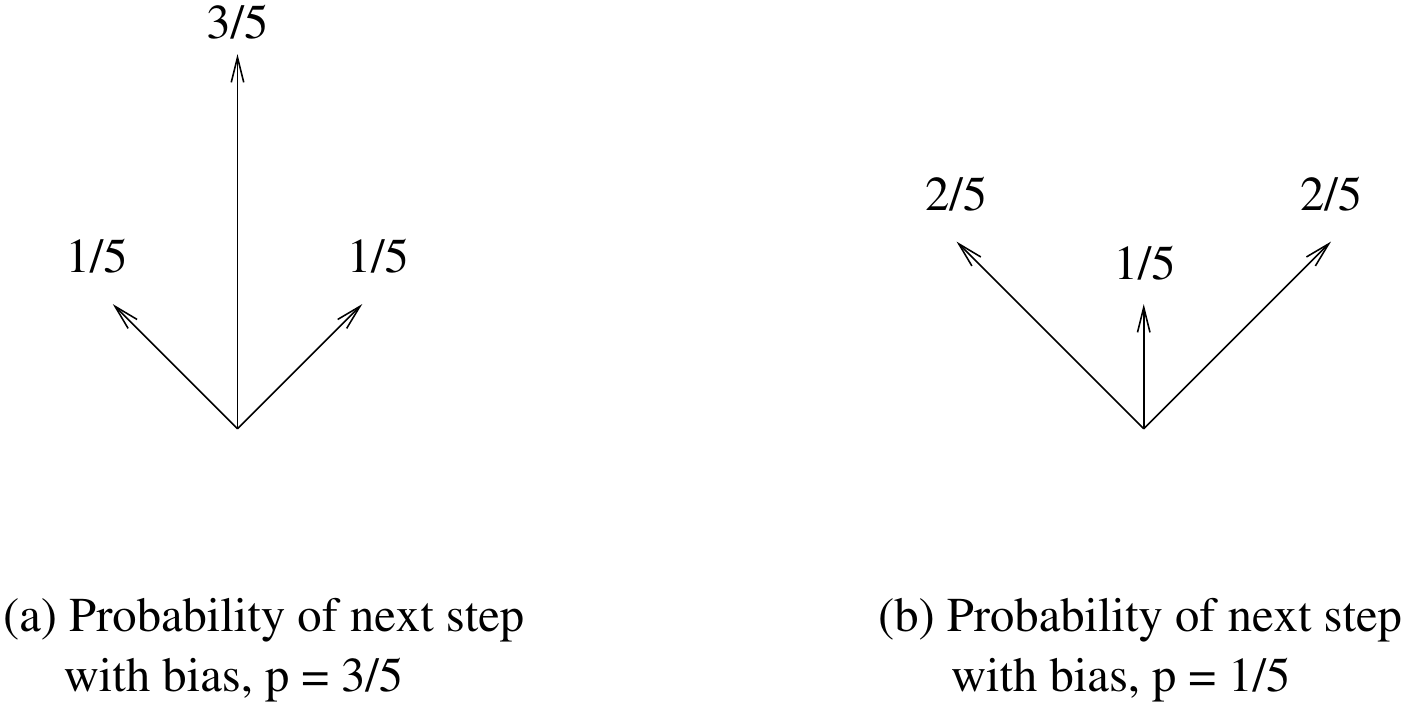}
  \end{center}
  \caption{Directionally biased walk for agent with current direction north and biases $p=3/5$ and $p=1/5$.}
   \label{fig:db}
\end{figure}

We do not adopt the von Mises distribution in our approach for two reasons. Firstly, we have only a discrete number of directions and so do not require a continuous distribution. Secondly, as in random walks, we would like the computations the agent needs to perform to be simple. Our notion of directional bias limits our agent to choose either its current direction with a probability $p$ (referred to as the {\em bias\/}), or any neighbouring direction, i.e., $\pi/4$ radians ($45^o$) clockwise or anti-clockwise from the current direction, with equal probability of $(1-p)/2$. When the bias $p$ is high (as illustrated in Fig.~\ref{fig:db}(a)), the movement model approximates (discretely) that of the von Mises distribution for a high value of $\kappa$ (such as $\kappa=4$ in Fig.~\ref{fig:vm}). This is not the case, however, for low values of $p$ (as illustrated in Fig.~\ref{fig:db}(b)).

Let $Direction = \{0, \pi/4, \pi/2, 3\pi/4, \pi, 5\pi/4, 3\pi/2, 7\pi/4\}$ be the possible directions in radians. Our notion of directional bias is then defined formally as follows.

\vspace{1mm}
\begin{definition} (Directional bias) 
\label{def:bias}
Given the current direction $d\mem Direction$ and bias $p \mem [0,1]$, the probability of moving in direction $d'\mem Direction$ at the next step, $P(d')$, is defined as follows.

\[d'=d \implies P(d')=p\\
d'\mem \{(2\pi+ (d-\pi/4))\mod 2\pi, (d+\pi/4)\mod 2\pi\} \implies  P(d')=(1-p)/2\\
d'\nmem \{d,(2\pi+(d-\pi/4)) \mod 2\pi, (d+\pi/4)\mod 2\pi\} \implies P(d')=0 \t1~~~ \Box\]
\end{definition}

We also investigate adding occasional random steps to our directionally biased walks. The idea is that with probability $r$ the agent will make a random, rather than directionally biased, step. This better matches our own observations of the movements of ants. 

\vspace{1mm}
\begin{definition} (Directional bias with random steps) 
\label{def:r}
Given the current direction $d\mem Direction$, bias $p \mem [0,1]$ and probability $r\mem [0,1]$ of  a random step, the probability of moving in direction $d'\mem Direction$ at the next step, $P(d')$, is defined as follows.

\[d'=d \implies P(d')=r*1/8 + (1-r)*p\\
d'\mem \{(2\pi+(d-\pi/4)) \mod 2\pi, (d+\pi/4)\mod 2\pi\} \implies\\
\t1 P(d')=r*1/8 + (1-r) (1-p)/2\\
d'\nmem \{d,((2\pi+(d-\pi/4))\mod 2\pi, (d+\pi/4)\mod 2\pi\} \implies  P(d')=r*1/8 ~~ \Box\]
\end{definition}

To analyse coverage time under our models of directional bias, we adapt a Markov-chain model \cite{nor98} developed for random walks by Mian et al.\ \cite{mia10}. As explained in Section~\ref{sec:macro}, this allows us to calculate the coverage time directly, and hence compare coverage times for different network sizes and biases. 

\section{A macro-level model}
\label{sec:macro}

The previous work on directional bias by Fink et al.\ \cite{fin12} shows that directionally biased walks are more efficient than random walks on a regular, connected toroidal graph; but only for the one specific network size and bias considered in their paper. In this paper, we produce more general results by investigating the effect on coverage time of varying the network size and directional bias. The micro-level model and simulation approach used by Fink et al.\ is not suited to this goal, requiring numerous simulations runs to calculate the coverage time for each network size and bias. We therefore use a more abstract, macro-level model which allows us to calculate the coverage time for a given network size and bias directly. 

Our model is based on the work of Mian et al. \cite{mia10} who provide a Markov-chain approach \cite{nor98} to model and calculate coverage time for random walks on a regular, connected toroidal graph. Given a network of $N$ nodes, let the vector $v$ of length $N$ denote the state probability distribution with elements $v_i$ for $0 \leq i < N$, and the matrix $M$ of size $N\cross N$ denote the transition probability matrix with elements $M_{i,j}$ for $0 \leq i,j <N$. 

\vspace{1mm}
\begin{definition} (Markov-chain model) 
\label{def:markov}
Let $v^{(0)}$ denote the initial state distribution and $v^{(k)}$ denote the state distribution after the $k$th step. 

\begin{itemize}
\item The elements of the state probability distribution sum to 1. 

\[\all k \geq 0\dot  \,\underset{i=0}{\stackrel{N-1}{\sum}} v^{(k)}_i = 1\]

\item The rows of the transition probability matrix sum to 1. 

\[\all i < N \dot \,\underset{j=0}{\stackrel{N-1}{\sum}} M_{i,j} = 1\]

\item The state distribution at step $k$ is calculated by multiplying the initial distribution by the transition probability matrix $k$ times. 

\[\all k > 0\dot v^{(k)} = v^{(0)} M^{k} \t9\t1\, \Box\]
\end{itemize}
\end{definition}

Often the system described by such a model would begin in a particular node with probability 1. 

\[\exi i < N \dot v_i^{(0)} = 1 \land (\all j < N \dot j \neq i \implies v_j^{(0)}=0)\]
For a random walk, we call this node the {\em starting node\/}. A random walk is specified by letting the transition from a node to any of its neighbours occur with a probability of $1/n$ where $n$ is the number of neighbours. For example, for a 1-dimensional network of 5 nodes such that 

\[v^{(0)} = \begin{pmatrix}0~&  0~&  1~&  0 ~& 0\end{pmatrix}\]
the transition probability matrix for a random walk would be 

\[M=\begin{pmatrix}
0 ~& \frac{1}{2} ~& 0~ & 0~ & \frac{1}{2}~\\
\frac{1}{2}~ & 0~ & \frac{1}{2}~ & 0~ & 0~\\
0~ & \frac{1}{2}~ & 0~ & \frac{1}{2}~ & 0~\\
0~ & 0~ & \frac{1}{2}~ & 0~ & \frac{1}{2}~\\
\frac{1}{2}~ & 0~ & 0~ &\frac{1}{2}~ & 0~
\end{pmatrix}\]
where, for example, row 0 (the topmost row of $M$) indicates that an agent at node 0 (the first node of $v$) has a probability of $1/2$ of moving to node 1 and a probability of $1/2$ of moving to node 4 (since we wrap the east and west edges). Multiplying $v^{(0)}$ by $M$ results in

\[v^{(1)} = \begin{pmatrix}0 ~&\frac{1}{2}~& 0~& \frac{1}{2}~& 0\end{pmatrix}\]
\vspace{-8mm}
\[v^{(2)} =\begin{pmatrix}\frac{1}{4}~& 0~& \frac{1}{2}~& 0~& \frac{1}{4}\end{pmatrix}\]
and so on. For a 2-dimensional network of $n \cross m$ nodes, the vector $v$ would have $n*m$ elements, with those from $i*m\upto i*m+m-1$ for $0 \leq i < n$ denoting the nodes in the $i$th row of the network. So, for example, row 0 of the matrix for a random walk on a network of 5 $\cross$ 5 nodes would be

\[\underbrace{\begin{matrix} 0 ~& \frac{1}{8} ~& 0 ~& 0 ~& \frac{1}{8}\end{matrix}}_{row 0}~ \underbrace{\begin{matrix}\frac{1}{8} ~& \frac{1}{8} ~& 0 ~& 0 ~& \frac{1}{8}\end{matrix}}_{row 1}~ \underbrace{\begin{matrix} 0 ~& 0 ~& 0 ~& 0 ~& 0 \end{matrix}}_{row 2}~ \underbrace{\begin{matrix} 0 ~& 0 ~& 0 ~& 0 ~& 0\end{matrix}}_{row 3}~ \underbrace{\begin{matrix} \frac{1}{8} ~& \frac{1}{8} ~& 0 ~& 0 ~& \frac{1}{8}\end{matrix}}_{row 4}\]
where $row0$ ro $row4$ refer to the row in the network.

In order to calculate coverage time, we modify the standard Markov-chain model for a random walk so that the starting node is an {\em absorbing node\/}, i.e., a node from which the probability of a transition to any neighbour is 0 (and the probability of a transition to itself is 1). We then model the system as starting from the state distribution after the initial distribution, i.e., that in which all neighbours of the starting node have probability $1/n$ where $n$ is the number of neighbours per node. 

\vspace{1mm}
\begin{definition}(Markov-chain model with absorbing node)
\label{def:absorb}
Given a Markov-chain model as defined in Definition~\ref{def:markov}, let $s < N$ be the position of the starting node. A Markov-chain model with absorbing starting node is defined in terms of a transition probability matrix $M'$ and initial state probability distribution $v^{(0)}$ as follows.

\begin{itemize}
\item The initial state probability distribution is that reached after 1 step from a distribution in which the agent is in the starting node with probability 1.

\[v^{(0)} = v\, M \t1 {\rm where~} v_s=1 \land (\all j < N \dot  j\neq s \implies v_j=0)\]

\item The starting node transitions to itself with probability 1 in $M'$.

\[M'_{s,s} = 1 \land (\all j < N \dot j \neq s \implies M'_{s,j}=0)\]

\item All other transitions in $M'$ are as in $M$.

\[\all i, j < N \dot i\neq s \implies M'_{i,j}=M_{i,j} \t8~\Box\]
\end{itemize}
\end{definition}

Due to the starting node being an absorbing node, the probability of being in the starting node never decreases as the number of steps increase. For example, for the 1-dimensional network above, the transition probability matrix is

\[M'=\begin{pmatrix}
0 ~& \frac{1}{2} ~& 0~ & 0~ & \frac{1}{2}~\\
\frac{1}{2}~ & 0~ & \frac{1}{2}~ & 0~ & 0~\\
0~ & 0~ & 1~ & 0~ & 0~\\
0~ & 0~ & \frac{1}{2}~ & 0~ & \frac{1}{2}~\\
\frac{1}{2}~ & 0~ & 0~ &\frac{1}{2}~ & 0~
\end{pmatrix}\]
and the state probability distribution is

\[v^{(0)} = \begin{pmatrix}0~& \frac{1}{2}~& 0~& \frac{1}{2}~& 0\end{pmatrix}\]
\vspace{-8mm}
\[v^{(1)} =\begin{pmatrix}\frac{1}{4}~& 0~& \frac{1}{2}~& 0~& \frac{1}{4}\end{pmatrix}\]
\vspace{-8mm}
\[v^{(2)} =\begin{pmatrix}\frac{1}{8}~& \frac{1}{8}~& \frac{1}{2}~& \frac{1}{8}~& \frac{1}{8}\end{pmatrix}\]
\vspace{-8mm}
\[v^{(3)} =\begin{pmatrix}\frac{1}{8}~& \frac{1}{16}~& \frac{5}{8}~& \frac{1}{16}~& \frac{1}{8}\end{pmatrix}\]
\vspace{-8mm}
\[v^{(4)} =\begin{pmatrix}\frac{3}{32}~& \frac{1}{16}~& \frac{11}{16}~& \frac{1}{16}~& \frac{3}{32}\end{pmatrix}\]
and so on. 

The probability of the starting node at step $k$ in this model is the probability that the system has returned to the starting node within $k$ steps. It can be used to calculate the coverage time as follows.

Let $\gamma_i$ denote the expected number of {\em new\/} nodes covered in the $i$th step. The total expected number of nodes covered at the $k$th step, $C_k$, is then 

\[C_k = \,\underset{i=0}{\stackrel{k}{\sum}} \gamma_i\,.\]
The initial node is covered at step 0, so we have $\gamma_0=1$. For all $k > 0$, $\gamma_k$ is equal to the probability that the node, $n_k$, reached at the $k$th step has not been visited before, i.e., $\gamma_k = P(n_k\nmem \{n_i | i < k\})$ or 

\[\gamma_k=P(n_k\neq n_0 \land \ldots \land n_k\neq n_k-1)\,.\]
Due to the regularity of the network and the fact that an agent behaves the same at each node, the probability of returning to the starting node after, say, 10 steps is equal to the probability of returning to the second node reached after 12 steps. More generally, we have $P(n_{k-i}=n_0) = P(n_k= n_i)$. From which it follows that $P(n_{k-i}\neq n_0) = P(n_k\neq n_i)$. Hence, from above

\[\gamma_k = P(n_0\neq n_1 \land \ldots \land n_0\neq n_k)\]
which is equal to the probability that the system has not returned to the starting node, $n_0$, within $k$ steps. In other words, given the modified Markov-chain model of Definition~\ref{def:absorb}

\[C_k = \,\underset{i=0}{\stackrel{k}{\sum}} (1-v^{(i)}_s)\,.\]
Coverage time can then be defined as follows. 

\vspace{1mm}
\begin{definition}(Coverage time)
\label{def:coverage}
Given a Markov-chain model with absorbing node as defined in Definition~\ref{def:absorb}, the time to cover $x$\% of the nodes is the smallest $k \geq 0$ such that 

\[C_k = \,\underset{i=0}{\stackrel{k}{\sum}} (1-v^{(i)}_s)=(x/100)*N\,. \t8~~~\Box\]
\end{definition}

With directional bias, we add an additional dimension to our representation of a network: the current direction of movement. For a network with $N$ nodes, the number of entries in the state probability distribution $v$ is hence no longer $N$ but $n*N$ where $n$ is the number of neighbours per node (and hence the number of directions of movement). Each entry represents the probability of being in a node having entered from a specific direction. We organise these entries so that those from $d*N\upto d*N+N-1$ for $0 \leq d < n$ denote the probabilities of being in a node of the network having entered from direction $d$. The corresponding  transition probability matrix $M'$ for the Markov-chain model is of size $n*N \cross n*N$, and since there are $n$ positions corresponding to the starting node (one for each direction from which the starting node was entered) there will be $n$ absorbing positions in the matrix. 

\vspace{1mm}
\begin{definition} (Markov-chain model for directional bias)
\label{def:mdb}
Let $M$ be a transition probability matrix of size $n*N\cross n*N$ (such that all rows sum to 1). A Markov-chain model for directional bias is defined in terms of a transition probability matrix $M'$ and initial state probability distribution $v^{(0)}$ as follows.

\begin{itemize}
\item The initial state probability distribution is that reached after 1 step from a distribution in which the agent is in the starting node (with a particular current direction) with probability 1.

\[v^{(0)} = v\, M \\
{\rm where~} \M\exi d < n \dot v_{d*N+s}=1 \land (\all j < n*N \dot j \neq d*N+s \implies v_j=0)\O\]

\item The starting node transitions to itself (without changing the current direction) with probability 1 in $M'$.

\[\all d < n \dot \M M'_{d*N + s,d*N+s} = 1 \land\\
 (\all j <  n*N\dot j \neq d*N+s \implies M'_{d*N+s,j}=0)\O\]

\item All other transitions in $M'$ are as in $M$.

\[\all i, j < n*N \dot \all d < n \dot i\neq d*N+s \implies M'_{i,j}=M_{i,j}  \t3~~\Box\]
\end{itemize}
\end{definition}

For example, consider adding directional bias to the 1-dimensional network of 5 nodes. Let the probability of continuing in the current direction be $p=3/4$, and that of changing direction to be $(1-p)=1/4$. Let 0 denote direction east (or right) and 1 denote direction west (or left), and let the starting node be node 2 (the centre node). The transition probability matrix is then

\begin{sidebyside}
\[M'= \begin{pmatrix} 
0 ~& \frac{3}{4} ~& 0 ~& 0 ~& 0 ~&    0 ~& 0 ~& 0 ~& 0 ~& \frac{1}{4} \\
0 ~& 0 ~& \frac{3}{4} ~& 0 ~& 0 ~&    \frac{1}{4} ~& 0 ~& 0 ~& 0 ~& 0 \\
0 ~& 0 ~& 1 ~& 0 ~& 0 ~&    		   0 ~& 0 ~& 0 ~& 0 ~& 0 \\
0 ~& 0 ~& 0 ~& 0 ~& \frac{3}{4} ~&     0 ~&  0 ~& \frac{1}{4} ~& 0 ~& 0 \\
\frac{3}{4} ~& 0 ~& 0 ~& 0 ~&  0~&     0 ~& 0 ~&  0 ~& \frac{1}{4} ~& 0 \\
0 ~& \frac{1}{4} ~& 0 ~& 0 ~& 0 ~&    0  ~& 0 ~& 0 ~& 0 ~& \frac{3}{4}\\
 0 ~& 0 ~& \frac{1}{4} ~& 0 ~& 0 ~&   \frac{3}{4} ~& 0~& 0 ~&  0 ~& 0 \\  
0 ~& 0 ~& 0 ~& 0 ~& 0 ~&    		    0 ~& 0 ~& 1 ~& 0 ~& 0 \\
0 ~&  0 ~&  0 ~& 0 ~& \frac{1}{4} ~&   0 ~& 0 ~& \frac{3}{4} ~& 0 ~& 0 \\
\frac{1}{4} ~& 0 ~& 0 ~& 0 ~&   0 ~&    0 ~& 0 ~& 0 ~& \frac{3}{4} ~&  0
\end{pmatrix} \\
\t3~~\underbrace{\t5~~}_{east} \underbrace{\t5~~}_{west}\]
\nextside
\[\hspace*{5mm}\left. \begin{aligned} ~\\~\\~\\~\\ \end{aligned}\right\} ~ \mbox{\scriptsize $east$}\]  
\vspace{1mm}
\[\hspace*{5mm}\left. \begin{aligned} ~\\~\\~\\~\\ \end{aligned}\right\} ~ \mbox{\scriptsize $west$}\]
\end{sidebyside}
where, for example, row 0 indicates that an agent at node 0 whose current direction is east will move to node 1 (and not change the direction) with probability $3/4$ and to node 4 (changing the current direction to west) with probability $1/4$. Also, rows 2 and 7 indicate that an agent in node 2 will remain in node 2 and not change the current direction. 

If the starting node's initial direction is set to east then the state probability distribution is 
 
\[v=\begin{pmatrix} 0~& 0~& 1~&  0~&  0~&  0~&  0~&  0~&  0~&  0\end{pmatrix}\]
\vspace{-14mm}
\[\t1~\underbrace{\t2~~~}_{east} \,\underbrace{\t2~~~}_{west}\]
and hence the initial state probability distribution for the Markov-chain is 

\[v^{(0)} = \begin{pmatrix} 0~& 0~& 0~&  \frac{3}{4}~&  0~&  0~&  \frac{1}{4}~&  0~&  0~&  0\end{pmatrix}\]
\vspace{-14mm}
\[\t1~~~~\,\underbrace{\t2~~~}_{east} \ \,\underbrace{\t2~~~}_{west}\]

Given that all nodes in our network follow the same directional-bias rules, we can calculate coverage time in a manner similar to that in Definition~\ref{def:coverage}. The difference is that we need to sum the probabilities from the positions in the state probability matrix $M'$ corresponding to the starting node. 

\vspace{1mm}
\begin{definition}(Coverage time for directional bias)
\label{def:cdb}
Given a Markov-chain model for directional bias as defined in Definition~\ref{def:mdb}, the time to cover $x$\% of the nodes is the smallest $k \geq 0$ such that 

\[C_k = \,\underset{i=0}{\stackrel{k}{\sum}} (1- \,\underset{d=0}{\stackrel{n-1}{\sum}} v_{d*N+s}^{(i)})=(x/100)*N\,. \t6~~~~~\Box\]
\end{definition}

\section{Investigating directional bias}
\label{sec:res}

To perform our investigation into directional bias, we implemented our Markov-chain model for directional bias (Definition~\ref{def:mdb}) and coverage time formula (Definition~\ref{def:cdb}) in Java using the JAMA library for matrices and matrix operations ({\tt math.nist.gov/javanumerics/jama}) and the JFreeChart library for plotting graphs ({\tt www.jfree.org/freechart}). 

Initially, we plotted graphs of coverage time versus bias (for bias values from 0 to 0.95 in steps of 0.05) for graphs sizes $5\cross 5$ (25 nodes) to $15 \cross 15$ (225 nodes). We calculated the time for coverage of 99\% of the network nodes. This was to avoid problems arising with 100\% coverage when the coverage converged to a point just below the network size due to inaccuracies in the floating-point arithmetic.

\begin{figure*}[t]
\centering
 \hspace*{-1mm}\includegraphics[scale=0.285]{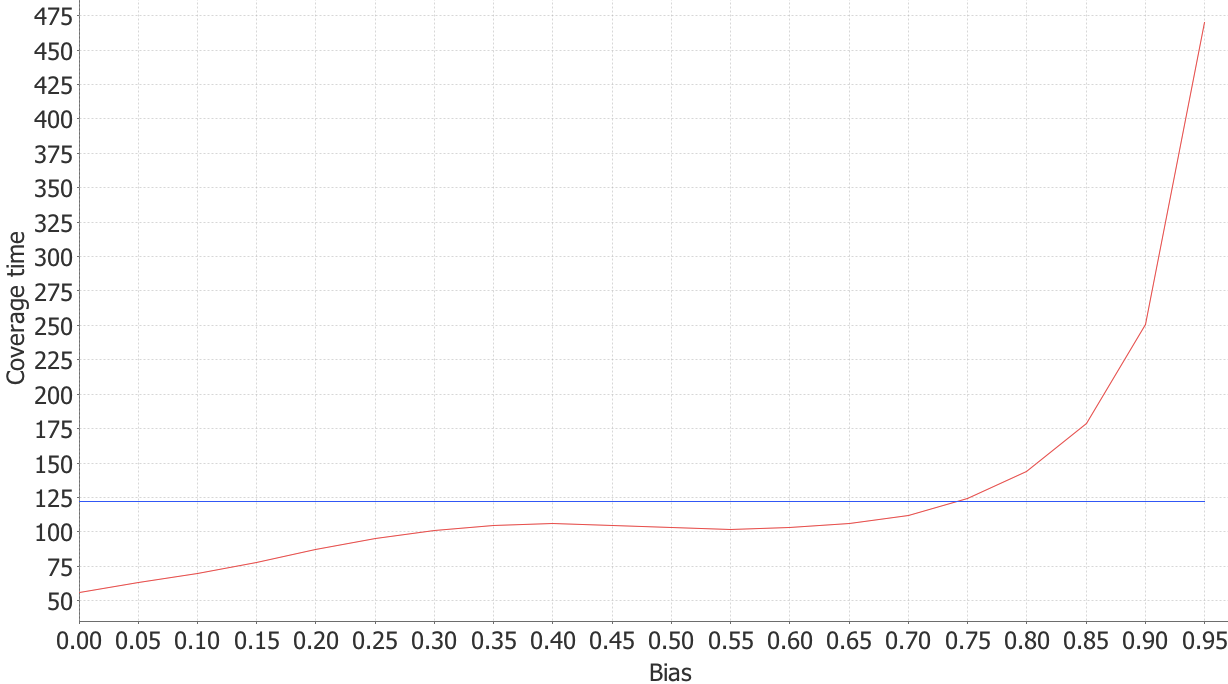}
  \caption{Random vs directionally biased walk for a 2-dimensional network of 5 $\cross$ 5 nodes.}
   \label{fig:55}
\end{figure*}

\begin{figure*}[t]
\centering
 \hspace*{-1mm}\includegraphics[scale=0.285]{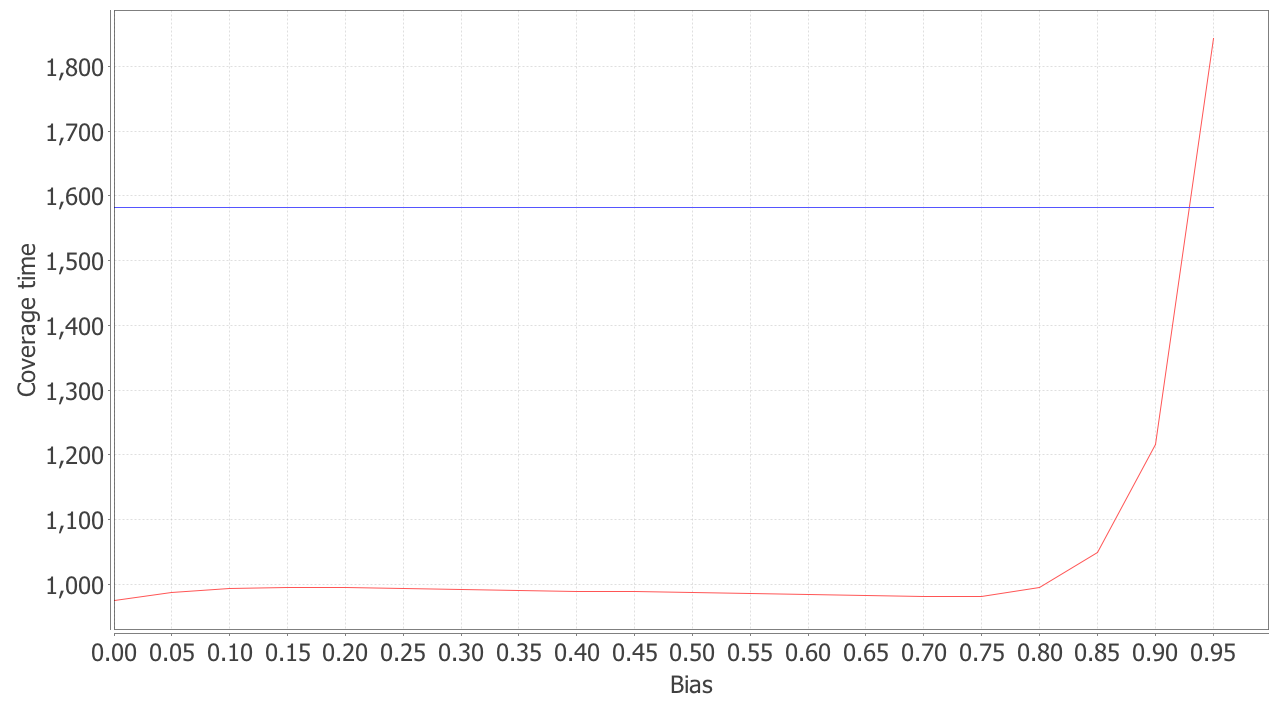}
  \caption{Random vs directionally biased walk for a 2-dimensional network of 15 $\cross$ 15 nodes.}
   \label{fig:1515}
\end{figure*}

The movement model was that of Definition~\ref{def:bias}. The graphs for the $5\cross 5$ and $15\cross 15$ networks are shown in Fig.~\ref{fig:55} and Fig.~\ref{fig:1515}, respectively. The horizontal line represents the coverage time for a random walk, and the curved line that for a directionally biased walk under the range of biases. The general shape of the latter and its position in relation to the horizontal line for random bias was consistent for all network sizes in the range considered.

A number of interesting results follow from this analysis.

 \begin{enumerate}
\item The best coverage time is achieved for a bias of 0. This corresponds to an agent which always changes direction by $\pi/4$ radians on every step. 
\item While for low directional biases (0 up to around 0.7 for the $5 \cross 5$ case) coverage time is less than that for a random walk, for higher biases it is greater than that for a random walk. 
\item The value of the bias at which a directionally biased walk becomes less efficient than a random walk (from here on called the {\em cross-over bias\/}), progressively increases as the size of the network increases. It is around 0.74 for a $5\cross 5$ network, and 0.93 for a $15\cross 15$ network. 
\item The improvement in efficiency of directional bias increases as the size of the network increases. For a directional bias of 0.5 the increase in efficiency is less than 20\% for a $5\cross 5$ network, and around 40\% for a $15\cross 15$ network.
\end{enumerate}

Point~1 is particularly interesting as it suggests a new movement model that was not initially anticipated. Our initial motivation was to investigate movement models similar to those observed in nature, which are best represented by a von Mises distribution. As illustrated in Fig.~\ref{fig:db}(b), however, low values of bias in our movement model (including the value 0) do not correspond to a von Mises distribution. The new model, although perhaps impractical as a means of movement in nature, can nevertheless be readily implemented in network search and dissemination algorithms. 

Point~2 is also interesting as it indicates that directional bias is only effective in reducing coverage time when the bias is not too large. This result was also unanticipated as directional bias in the movement of animals tends to be high. However, the areas over which such animals move would correspond to networks significantly larger than those we considered. Point~3 anticipates that the cross-over bias would be higher in such networks. This conjecture is supported by the work of Fink et al.\ \cite{fin12} whose micro-level simulation of a 2-dimensional network of 100 $\cross$ 100 nodes shows that a directionally biased walk (approximating a von Mises distribution with high $\kappa$) is more efficient than a random walk. 

\begin{figure*}[t]
\centering
 \hspace*{-1mm}\includegraphics[scale=0.285]{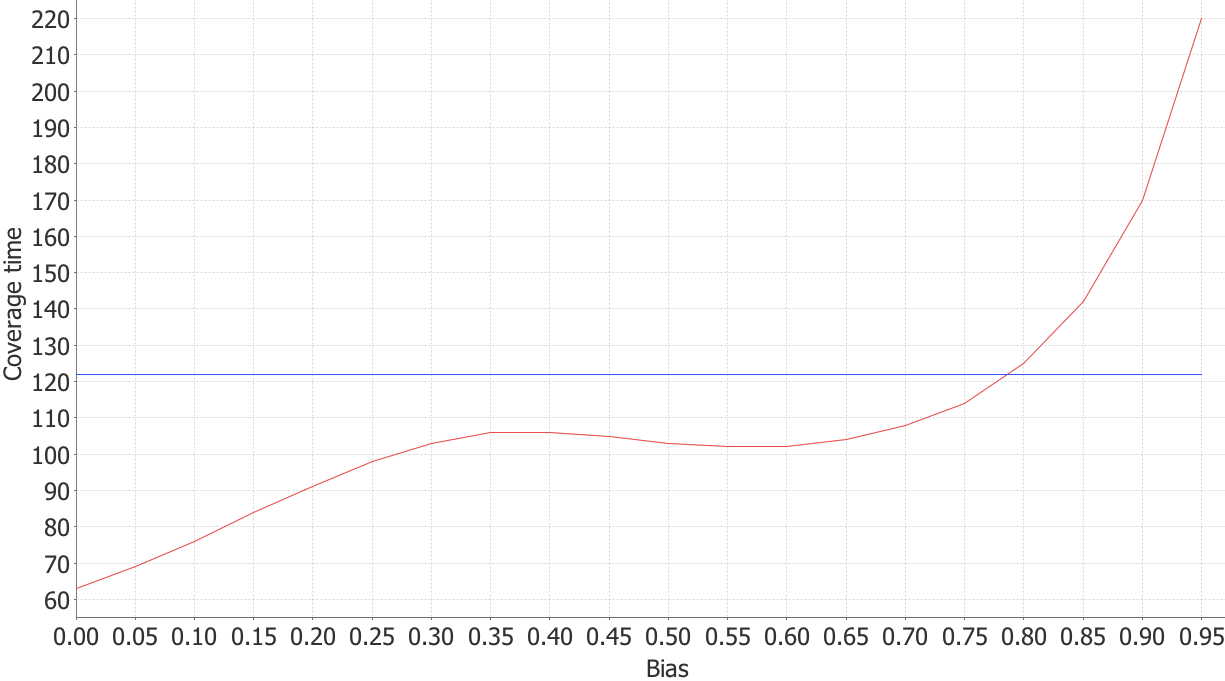}
  \caption{Random vs directionally biased walk with probability 0.1 of a random step for a 2-dimensional network of 5 $\cross$ 5 nodes.}
   \label{fig:r10}
\end{figure*}

\begin{figure*}[t]
\centering
 \hspace*{-1mm}\includegraphics[scale=0.285]{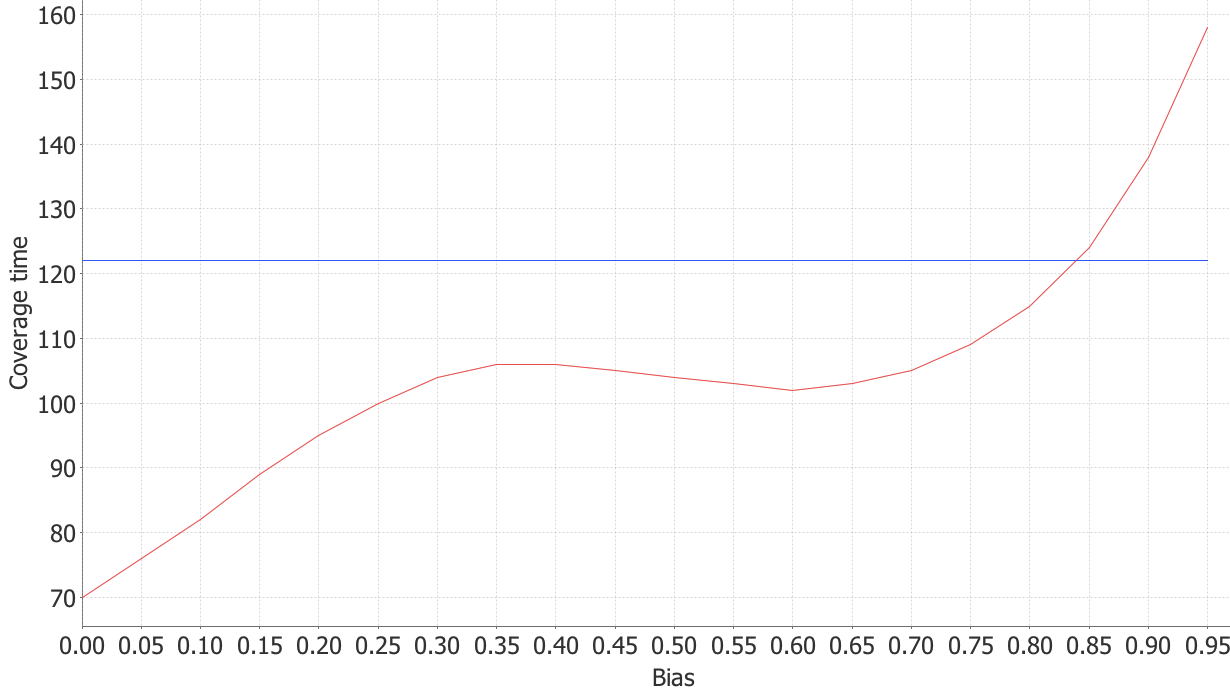}
  \caption{Random vs directionally biased walk with probability 0.2 of a random step for a 2-dimensional network of 5 $\cross$ 5 nodes.}
   \label{fig:r20}
\end{figure*}

The second part of our investigation considered the movement model of Definition~\ref{def:r}, where occasional random steps are added to a directionally biased walk. Example plots for a network of $5\cross 5$ nodes and the value of $r$ set to 0.1 (an average of one random step in every 10) and 0.2 (an average of one random step in every 5) are shown in Fig.~\ref{fig:r10} and Fig.~\ref{fig:r20}, respectively. 

The following results emerge from this  analysis.

\begin{enumerate}
\item As may have been predicted, the addition of random steps moves the coverage time closer to that or a random walk. Hence, for bias values lower than the cross-over bias the coverage time increases, but for values higher than the cross-over value the coverage time decreases. 
Comparing Fig.~\ref{fig:r20} with Fig.~\ref{fig:55} it can be seen that for a bias of 0 the coverage time has increased from around 55 to 70, and for a bias of 0.95 it has decreased significantly from around 470 to about 158. 
\item The introduction of random steps increases the cross-over bias. For $r=0.1$ the cross-over bias increases to 0.78 (from 0.74 for no random steps) and for $r=0.2$ to 0.84. 
\end{enumerate}

\section{Conclusion}
\label{sec:conc}

In this paper we have investigated the effect of directional bias on the coverage time of random walks on regular, connected networks. Our analysis has shown that directional bias can reduce coverage time significantly and has a greater effect the larger the network. However, this reduction occurs only when the bias (to continue in the same direction) is below a certain value we call the {\em cross-over bias\/}. The cross-over bias is dependent on the network size, increasing as the size of the network increases. Hence, high values of bias which work well at reducing coverage time in large networks, may be less effective, or even increase the coverage time, in smaller networks. 

The cross-over bias can be increased by adding occasional random steps to a directionally biased walk --- the more random steps, the higher the cross-over bias. Adding such steps, however, moves the coverage time of a directionally biased walk closer to that of a random walk (increasing coverage time for biases below the cross-over bias).

Our analysis also revealed that a movement model in which an agent changes direction by $\pi/4$ radians (in either direction with equal probability) on each step is more effective in reducing coverage time than a standard directionally biased model. Further investigation of this, and similar models, is warranted. 

Our investigation was facilitated by a macro-level model of random and directionally biased walks in terms of Markov chains. This model allowed us to calculate coverage time directly, in contrast to other approaches where coverage time is calculated as the average result obtained from numerous runs of a micro-level simulation. Although introducing a margin of error due to the limitations of floating-point arithmetic, our more abstract model has provided a practical means for obtaining a deeper, more complete analysis of directional bias than in previous work.

\subsection*{\bf Acknowledgements}
This work was supported by Australian Research Council (ARC) Discovery Grant DP110101211.

\bibliographystyle{plain}
\bibliography{ants}

\end{document}